\documentclass[conference]{IEEEtran}
\IEEEoverridecommandlockouts
\usepackage{cite}
\usepackage{amsmath,amssymb,amsfonts}
\usepackage{algorithmic}
\usepackage{graphicx}
\usepackage{textcomp}
\usepackage{xcolor}
\usepackage[]{siunitx}
\usepackage{makecell}
\usepackage{booktabs}
\def\BibTeX{{\rm B\kern-.05em{\sc i\kern-.025em b}\kern-.08em
    T\kern-.1667em\lower.7ex\hbox{E}\kern-.125emX}}
\newcommand{\riscv}{\mbox{RISC-V}}
\begin{document}
\bstctlcite{IEEEexample:BSTcontrol}
\renewcommand{\arraystretch}{1.2}

\title{On-Demand Redundancy Grouping: Selectable Soft-Error Tolerance for a Multicore Cluster
\thanks{We acknowledge support by the EU H2020 Fractal project funded by ECSEL-JU grant agreement \#877056.}
}

\author{
\IEEEauthorblockN{Michael Rogenmoser\IEEEauthorrefmark{1}, Nils Wistoff\IEEEauthorrefmark{1}, Pirmin Vogel\IEEEauthorrefmark{2}, Frank Gürkaynak\IEEEauthorrefmark{1}, Luca Benini\IEEEauthorrefmark{1}\IEEEauthorrefmark{3}}
\IEEEauthorblockA{\IEEEauthorrefmark{1}\textit{ETH Zurich}, Zurich, Switzerland}
\IEEEauthorblockA{\IEEEauthorrefmark{2}\textit{lowRISC C.I.C.}, Cambridge, United Kingdom}
\IEEEauthorblockA{\IEEEauthorrefmark{3}\textit{University of Bologna}, Bologna, Italy}
}












\maketitle

\begin{abstract}
With the shrinking of technology nodes and the use of parallel processor clusters in hostile and critical environments, such as space, run-time faults caused by radiation are a serious cross-cutting concern, also impacting architectural design. This paper introduces an architectural approach to run-time configurable soft-error tolerance at the core level, augmenting a six-core open-source \riscv{} cluster with a novel On-Demand Redundancy Grouping (ODRG) scheme. ODRG allows the cluster to operate either as two fault-tolerant cores, or six individual cores for high-performance, with limited overhead to switch between these modes during run-time. 
The ODRG unit adds less than 11\% of a core’s area for a three-core group, or a total of 1\% of the cluster area, and shows negligible timing increase, which compares favorably to a commercial state-of-the-art implementation, and is 2.5$\times$ faster in fault recovery re-synchronization. Furthermore, when redundancy is not necessary, the ODRG approach allows the redundant cores to be used for independent computation, allowing up to 2.96$\times$ increase in performance for selected applications.

\end{abstract}


\section{Introduction}
In hostile environments such as space, radiation is ever-present and can have adverse effects on integrated circuits.
Particle strikes due to radiation can cause soft errors in combinational and sequential elements, where a resulting Single Event Upset (SEU) in a state-saving element. 
Traditionally, SEUs have impacted sequential elements such as memories and flip-flops and resulted in bit-flips in stored information. With the advent of smaller technology nodes and higher clock frequencies, transient errors that initially affect combinational logic and subsequently end up altering the state within sequential elements started to become a major concern~\cite{di_mascio_open-source_2021}.


In recent years, advances in computing power have enabled new applications, such as machine learning. In space, these new algorithms can be leveraged, for example to pre-process satellite images~\cite{di_mascio_leveraging_2019}, however adaptation is more difficult, as most processors for space lack the required computing power. When investigating the different application scenarios more closely, however, certain patterns appear: While some processing requires high levels of redundancy to ensure correct operation, other tasks may not require the full fault tolerance. As a more concrete example, a satellite will require reliability and correctness when processing commands and telemetry, however not for the imprecise processing of noisy data or running a machine-learning algorithm over an image, which could greatly benefit from increased processing power~\cite{furano_ai_2020}. 



In this work, we tackle this variability in application requirements by leveraging a run-time architectural configuration approach. We introduce \emph{On-Demand Redundancy Grouping (ODRG)}, an augmentation to a shared memory multi-core cluster architecture, which allows \riscv{} cores to be grouped on-demand, enabling Triple Modular Redundancy (TMR) at the core level. ODRG allows a six-core cluster to switch between operating as six individual cores for higher processing power and operating as two cores backed by TMR for critical tasks requiring fault tolerance. 

\section{Related Work} \label{sec:related}

As reliable processing is an important requirement for critical missions in space, several commercial processors that focus specifically on radiation tolerance have been developed. These radiation-hardened processors, such as the RAD line of processors~\cite{berger_quad-core_2015} implementing the PowerPC Instruction Set Architecture (ISA) or the LEON line of processors~\cite{andersson_leon_2017} implementing the SPARC ISA, are designed using radiation-hardened processes, leveraging characteristics of the fabrication and cell design that increase tolerance to SEUs, as well as ECC and modular redundancy for critical sequential elements in the design. Architectural redundancy mechanisms, such as ODRG presented in this work, can leverage standard commercial technologies in contrast to the specialized space-grade technologies, both reducing cost and improving performance for processors in space.

In recent years, the open-source \riscv{} ISA has gained significantly in popularity, leading to its consideration for processors for space~\cite{di_mascio_leveraging_2019, di_mascio_open-source_2021, gomez_-risc_2020}. 
The continuing development of the \riscv{} ISA can allow processors for space to leverage developments in efficiency and improvements in design, with its modularity targeting specialized application.


In~\cite{iturbe_arm_2019}, ARM presents a complete implementation to add fault tolerance within a core with their Triple Core Lock-Step (TCLS) architecture, a redundancy mechanism for their Cortex-R5 processor cores. To protect the execution of a single processor core, this core is triplicated, adding a TCLS Assist Unit as a wrapper around the three cores. With this TMR approach at the core level, all cores are connected to identical inputs, and the outputs are passed into a majority voter. Operating in lock-step, the outputs of the individual cores will be identical unless a fault corrupts the internal state of one of the cores. In case an error occurs within one of the cores, the voter will detect the fault if it reaches the core's interface and the TCLS Assist Unit will correct it. 



\section{Architecture}\label{sec:arch}
\begin{figure}
    \centering
    \includegraphics[width=0.9\columnwidth]{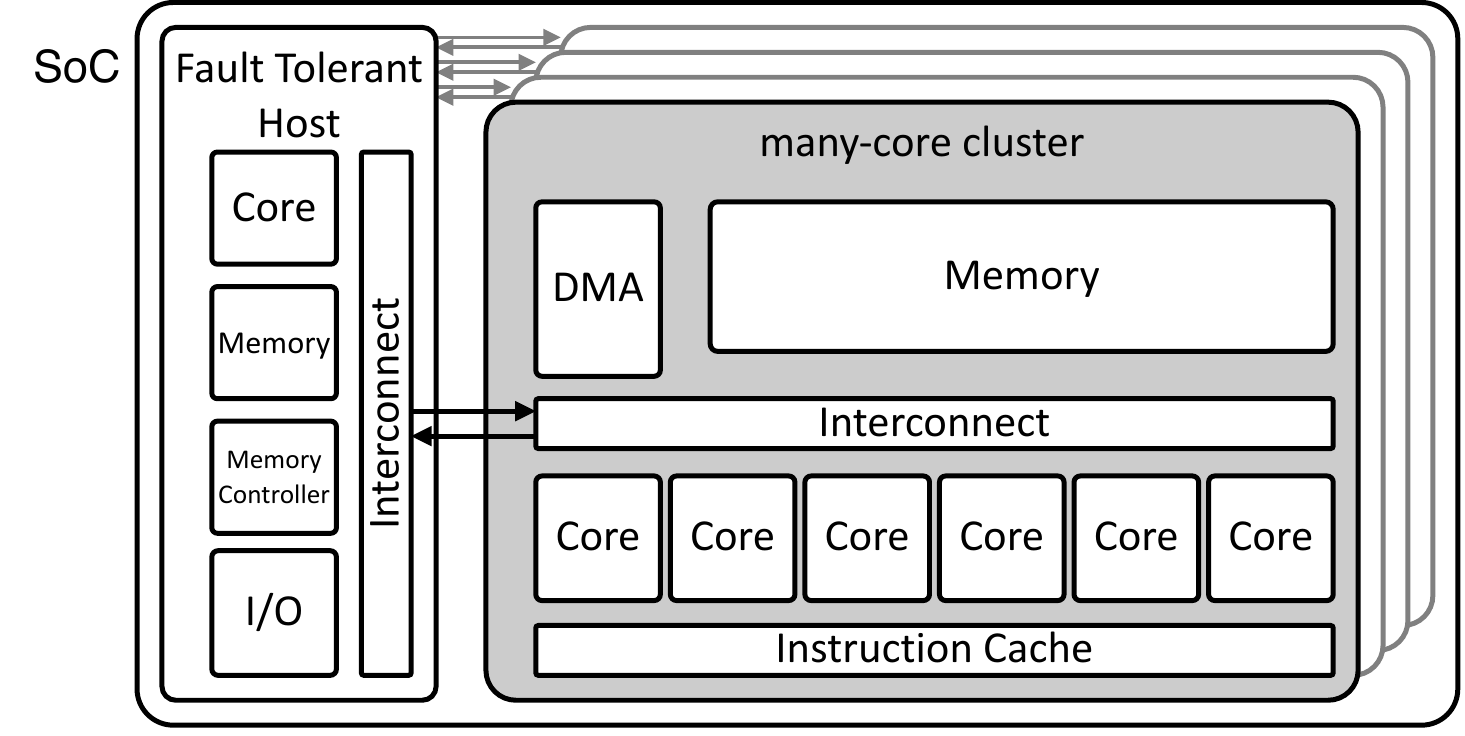}
    \caption{Overall architecture of an SoC with a general purpose host system supported by multi-core accelerators for data-parallel applications}
    \label{fig:system}
\end{figure}

In this work, we focus on a class of emerging computing architectures that consist of a powerful general-purpose host processor complemented by one or more accelerators based on multi-core clusters as seen in Figure~\ref{fig:system}. These clusters are designed to accelerate parallel tasks dispatched by a host processor. In a fault tolerance scenario, we will assume the host processor and peripherals are completely fault tolerant, protected by other means.

We focus on implementing a configurable redundancy mechanism in the multi-core cluster, accelerating computation tasks for critical applications requiring fault tolerance, as well as non-critical tasks that benefit from increased parallelism. As a baseline for this work, we use the open-source cluster from the Parallel Ultra-Low Power (PULP) platform~\cite{rossi_pulp_2015}. 

\subsection{Parallel Ultra-Low Power (PULP) cluster}

The PULP cluster contains multiple 32-bit \riscv{} cores. In this work the cluster was configured to include a total of six ``Ibex'' \riscv{} cores~\cite{davide_schiavone_slow_2017}, sharing an instruction cache. 
The cores' data ports are demultiplexed, allowing each core to primarily interact with a tightly coupled data memory (TCDM) with single-cycle access latency, here configured with \SI{64}{\kibi\byte} of memory grouped into 16 word-interleaved SRAM banks, connected through a logarithmic interconnect implemented as a crossbar with round-robin priority. The cores also have direct access to an event unit to manage synchronization across the parallel cores, as well as a DMA and various other peripherals and elements connected to a peripheral interconnect.
The SoC, connected to the cluster through two AXI interfaces, encompasses a capable host processing system with a larger L2 memory, as well as main memory controllers and I/O.

\subsection{On-Demand Redundancy Grouping (ODRG)}
\begin{figure}
    \centering
    \includegraphics[width=0.8\columnwidth]{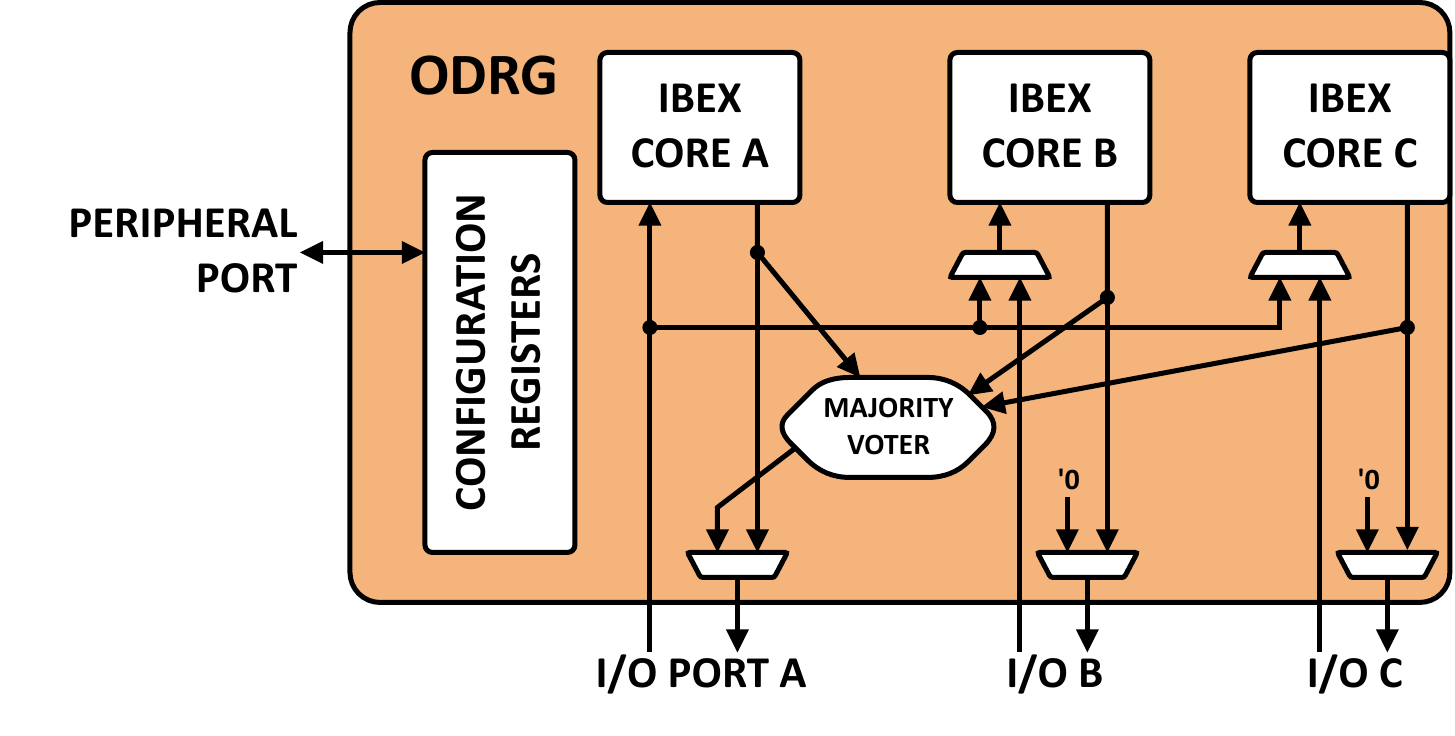}
    \caption{Block diagram of ODRG, wrapping 3 cores with a configurable voter}
    \label{fig:ODRG}
\end{figure}

The PULP cluster offers a unique advantage when designing a reliable architecture: Multiple cores are readily available, and a programming environment that supports the systems in various configurations has been developed independently. While usually the cluster cores are used in parallel to process different data more quickly, they can also be used to process identical data, which can subsequently be compared to detect errors in computing. ODRG leverages the multiple cores and adds architectural elements to directly allow three cores to operate on the same input, comparing the output using a majority voter. The ODRG architecture, shown in Figure~\ref{fig:ODRG}, groups together three cores within the cluster to one ODRG unit and can operate in the following two modes:

The \textit{performance} mode enables all cores to operate independently for maximum parallel performance. In this configuration, each core processes its own inputs and outputs, individually connecting to the instruction cache and the data interconnect. Each core is assigned its own hart-ID and has its own synchronization port within the event unit.

The \textit{soft-error tolerant} mode binds groups of three cores together into lock-step operation. In this configuration, all three cores receive identical inputs. The outputs of all cores are compared and voted upon to detect and correct any errors that may be present within the cores. The three cores operate as a single core, using only one interface to the rest of the system, which is otherwise occupied only by the first core in the group. The other two interfaces, typically occupied by the two cores now offering redundancy, are not used.

The majority voter in the ODRG is implemented individually for each bit, featuring additional detector circuitry comparing the output with the signals prior to voting to detect if a core has a fault. 
The number of faults per core are logged within the ODRG unit, allowing for software to distinguish between random faults, or a systematic issue within one core.

The two operational modes of an ODRG unit are implemented as states in a Finite State Machine (FSM).
To facilitate re-synchronization once an error is detected, the \textit{soft-error tolerant} mode in the FSM is split into three parts, one for normal operation, and one each to unload the internal state of the core and reload it back into the core.

In \textit{soft-error tolerant} mode, once a mismatch is detected, the ODRG unit will trigger a re-synchronization sequence, as a mismatch indicates differing internal states of the cores. This is done by switching the FSM into the unload state and signaling an interrupt to all cores in the ODRG group, which starts the re-synchronization software routine. This routine copies all internal registers onto the stack in memory, including relevant CSRs (such as the \texttt{MEPC} register). The stack pointer is saved to a register in the ODRG unit, signalling the unload is complete and giving all cores a reference to reload from. 
For this entire unload process, the majority voter is active, which allows any internal errors within one core to be corrected. 


In reload, the cores load the voted and stored state from memory back into the internal registers. As the stack pointer is saved at a known location, the state to reload is unique. Once the reload process is complete, the core returns to the PC stored prior to the interrupt, which due to synchronization of the \texttt{MEPC} register is identical for all three cores.




Within the PULP cluster, the ODRG unit exposes its configuration registers to the peripheral interconnect, allowing configuration from the host. These configuration registers are used to group and ungroup the cores, allowing the switch between \textit{soft-error tolerant} and \textit{performance} modes. Additional registers allow for the re-synchronization to be delayed, as well as store the stack pointer during re-synchronization. Finally, mismatches are counted for each of the cores and can be accessed through the same connection. 

\begin{figure}
    \centering
    \includegraphics[width=\columnwidth]{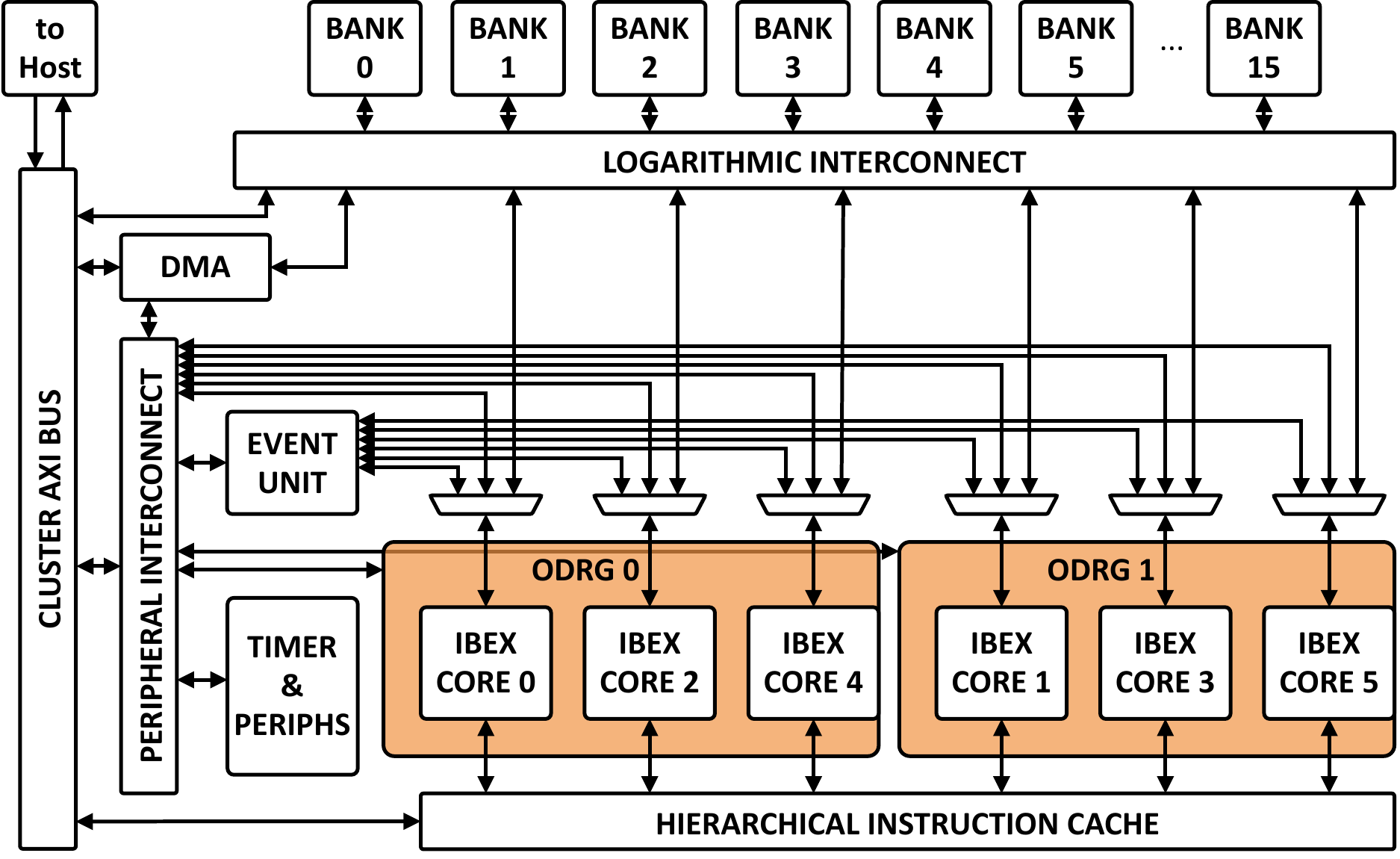}
    \caption{Block diagram of the PULP cluster with included ODRG units}
    \label{fig:odrg_cluster}
\end{figure}

In the implemented configuration shown in Figure~\ref{fig:odrg_cluster}, the six-core PULP cluster has two ODRG units, grouping three cores each.

\section{Evaluation}

\subsection{Setup}

To verify proper functionality and evaluate the performance impact ODRG has on the system, we simulate it at RTL level using \textsc{Siemens/Mentor Questasim v2019.3}. For these investigations, the cluster is integrated next to an Ibex host processor, additional memory, and peripherals within a host System-on-Chip (SoC). As our investigation concerns only the cluster, the Ibex host processor emulates a more capable host processing system and is responsible for configuring the ODRG unit and subsequently starting the cluster with the corresponding task. We use regression tests, which run highly parallel, performance-critical mathematical functions used in such as matrix-matrix multiplication and convolutions. 

\subsection{Performance}
The main advantage of our approach is the ability to trade-off performance with fault tolerance by activating ODRG only when needed. 
Table~\ref{tab:speedup} shows the performance improvement for three test applications that run in the ODRG \textit{soft-error tolerant} mode, as well as independently in \textit{performance} mode, with the corresponding parallelization across two or six cores. In a typical application scenario, the cores in the cluster will be used to accelerate parallelizable tasks to support the host system. For the highly parallelizable tasks simulated, we observe speedups between 2.63$\times$ and 2.96$\times$, which is close to the theoretical maximum of 3.00, showing an almost linear trade-off between redundancy and performance for tasks well suited to parallelization across six cores.  

\begin{table}[ht]
    \centering
    \caption{Performance of PULP cluster with ODRG in different modes}
    \begin{tabular}{@{}l rr r@{}}\toprule
        Test Application & \multicolumn{2}{c}{Cycle count} & Speedup \\
        \cmidrule{2-3}
         & \makecell{\textit{soft-error} \\ \textit{tolerant}} & \textit{performance} & \\ \midrule
        16-bit 2D Convolution & 100'938 & 36'379 & 2.77$\times$ \\
        32-bit 24x24 MatMul & 57'489 & 19'432 & 2.96$\times$\\
        32-bit 32x32 MatMul & 135'908 & 51'668 & 2.63$\times$\\
    \bottomrule
    \end{tabular}
    \label{tab:speedup}
\end{table}

While enabling re-configuring ODRG between \textit{soft-error tolerant} and \textit{performance} modes requires a reboot of the cluster, setting the mode requires very few cycles, as only a few memory-mapped registers are written. Rebooting and re-configuring the cluster takes less than 40'000 cycles, which equates to approximately \SI{160}{\micro\second} at \SI{250}{\mega\hertz}.
Therefore, switching between modes 100$\times$ per second would require less than 2\% overhead. Relating to specific applications, non-critical tasks which require more than 60'000 cycles in \textit{soft-error tolerant} mode can benefit from a speedup if 3$\times$ performance improvement is assumed. As typical applications, such as a machine learning algorithm, require many functions to execute in series, this overhead becomes negligible.

\subsection{Fault Injection}\label{sec:injection}

To evaluate the redundancy functions of ODRG, a limited fault injection analysis is performed in simulation. Using the force command within \textsc{Siemens/Mentor Questasim}, registers and signals are forced to an inverse value for a single cycle, simulating an SEU.

Fault injection is performed both directly at a core interface, forcing a signal exiting a core and entering the voter to the opposite value, as well as within a core's internal registers. As the interface is where errors are detected, a re-synchronization is immediately triggered. Injecting faults within a core's register file allows for observing how an internal corrupted state is handled. Once an error propagates to the core interface, a mismatch is detected and the re-synchronization is successfully able to correct the faulty register. As analyzed in more detail in~\cite{asciolla_characterization_2020}, not all faults within a core cause a noticeable failure of the system, as many of the errors are masked.

Triggering faults during the execution of a matrix multiplication show that the re-synchronization routine requires around 700~cycles to load the core state to memory through the voter, and back into the cores, roughly \SI{2.8}{\micro\second} at \SI{250}{\mega\hertz}. There is some variability of  $\pm 20$ cycles depending on the time of injection, due to the second grouped core continuing operation normally, leading to limited memory contention.

This compares favorably to the 1800-2400~cycles ARM reports are required for TCLS re-synchronization~\cite{iturbe_arm_2019}. Looking more closely, ARM stores 113~registers of the Cortex-R5 core, while only 41~registers are stored using Ibex.  In environments such as space, while SEUs occur far more frequently than on Earth's surface, they are still considered to be rare. On the other hand, environments such as particle accelerators~\cite{anelli_radiation_1999} observe far higher rates of SEUs, leading to a more stringent requirement on the re-synchronization time. In these contexts, the 2.5$\times$ re-synchronization speedup of ODRG with respect to TCLS may become a significant performance booster.

\subsection{Area Overhead}

To investigate the additional area required to support ODRG, we use \textsc{Synopsys DC 2019.12} to synthesize the PULP cluster in \textsc{GF} \SI{22}{\nano\meter} technology at \SI{250}{\mega\hertz} and worst-case conditions (SS, \SI{0.72}{\volt}, \SI{125}{\celsius}). A single ODRG adds around \SI{6.0}{\kilo GE}, and the total overhead of a cluster with two ODRG units and additional circuitry is \SI{25.6}{\kilo GE}. Compared to the base configuration with six Ibex cores and \SI{64}{\kibi\byte} ECC protected memory, ODRG adds less than 1\% to the total cluster area of \SI{2.16}{\mega GE}. An ODRG wrapping unit requires less than 11\% the area of a single Ibex core, which occupies \SI{60.5}{\kilo GE} in the implemented configuration.


The cluster both with and without ODRG was synthesized at a maximum clock frequency of \SI{415}{\mega\hertz} in worst-case conditions, with no discernible impact on the critical path.

This compares favorably to ARM's TCLS implementation, where they claim the TCLS unit adds around 4\% to the presented system, or around 14\% area of a Cortex R5 core~\cite{iturbe_arm_2019}. Furthermore, TCLS' clock frequency is approximately 10\% lower than a reference implementation, experiencing a significant hit due to the additional cores and voters required. Comparatively, ODRG adds less than 11\% of the 
Ibex core, with a negligible impact on timing.

\section{Discussion}

Safety is a crucial element required for a variety of processing tasks, especially in critical environments, but comes at a significant cost. With many implementations requiring \textgreater 3$\times$ the area to protect calculations with efficient correction, such as ARM TCLS~\cite{iturbe_arm_2019}, this cost can seem unreasonable for tasks and applications not requiring this level of redundancy.

In contrast to typical redundant architectures, ODRG leverages the additional cores included for redundancy to increase performance of non-redundant applications. Through simple re-configuration, these cores can operate in parallel, speeding up non-critical calculations and reducing the overall power requirements of the application. This can be especially valuable when using the same architecture in both ordinary environments as well as hostile environments, such as space, or when operating on already noisy data and executing functions that can tolerate minor deviations in these critical environments.

The ODRG approach is not specific to the core in use and no modifications were made to the Ibex \riscv{} cores used in this work. The ODRG method can be easily adapted to different processing cores and cluster configurations. 



\section{Conclusion}
In this work, we present On-Demand Redundancy Grouping (ODRG), a hardware architecture that enables the dynamic configuration of a compute cluster to run its cores either independently for high-performance, or redundantly for high reliability.
We implement ODRG on a six-core multi-core cluster based on the openly available PULP system~\cite{rossi_pulp_2015} and show that its \textit{performance} mode can speed up parallel tasks by up to 2.96$\times$, while reliably protecting against single event upsets (SEUs) in \textit{soft-error tolerant} mode.
We furthermore show that ODRG adds a minimal hardware overhead of less than \SI{1}{\percent} to the cluster and has little to no impact on frequency.

We find that ODRG is a flexible and low-cost solution for computing systems that are exposed to highly increased radiation levels, for instance when deployed in space missions.

\bibliographystyle{IEEEtran}
\bibliography{style,references}

\end{document}